\documentstyle[pra,floats,epsf,aps]{revtex}
\input epsf

\begin{document}
\draft
\title{Role of free energy landscape 
in the dynamics of mean field glassy systems}

\author{
Giulio Biroli
}
\address{Laboratoire de Physique Th{\'e}orique de l'Ecole Normale 
Sup{\'e}rieure \thanks{Unit{\'e} Mixte de Recherche du Centre Nationale de la 
Recherche Scientifique et de l'Ecole Normale Sup{\'e}rieure.},\\
24 rue Lhomond, 75231 Paris cedex 05, France}
\maketitle
\begin{abstract}
In this paper we expose the results of our recent work on the dynamical 
TAP approach to mean field glassy systems. Our aim is to
 clarify the connection between free energy landscape 
and out of equilibrium dynamics in solvable models.\\
Frequently qualitative explanations of glassy behaviour are based
on the ``Free Energy Landscape Paradigm''. If its relevance 
for equilibrium properties is clear, the relationship
between free energy landscape and out of equilibrium dynamics
is not well understood yet. In this paper we clarify
this relationship for the class of spin glass models which
reproduce phenomenologically some features of structural glasses. 
The method we use is a generalisation to dynamics of the 
Thouless, Anderson and Palmer approach to thermodynamics of
mean field spin glasses. Within this framework
we show to what extent the dynamics 
can be represented as an evolution in the free energy landscape.
In particular the relationship
between the long-time dynamics and the local properties
of the free energy landscape shows up explicitly using
this approach.
\end{abstract}

\pacs{PACS Numbers~:02.70.Ns, 61.20.Lc, 61.43.Fs. Preprint LPT-ENS 00/14}


\section{Introduction}
\label{sec:section1}
Generally in the study of thermodynamics much attention is payed
 to the free energy landscape \cite{landscape}.
 This landscape, which can be interpreted as the effective potential whose
 minima represent different possible states, 
gives an intuitive and quantitative description of the equilibrium
properties. Consider for example ferromagnetic systems. 
In this case the effective potential
 is a function of magnetisation. The ferromagnetic transition 
 corresponds to the splitting of the paramagnetic minimum in 
 the two ferromagnetic minima. At low temperature 
a vanishing external magnetic field breaks
 the up-down symmetry and fixes the system in one of the two possible
 ferromagnetic states.\\
Generally, glassy systems are characterised by a complicated energy
 landscape, which can give rise eventually to the existence of
 many possible states. Frequently, qualitative explanations of glassy
 behaviours are based on some assumptions on the properties of the 
free energy landscape \cite{landscape}.
Consider for example the Kirkpatrick-Thirumalai-Wolynes scenario
 for the glass transition \cite{vetri1,vetri2} 
in which the (exponential) number of 
states with a given free energy plays a crucial role.\\
However, if the relevance of the free energy landscape for the equilibrium
 properties is clear, the relationship between the free energy landscape and 
the dynamical behaviour is not completely understood, especially 
for glassy systems which remain
out of equilibrium also at long time. For instance
 what we can learn on the (out of equilibrium) dynamical behaviour starting
 from the knowledge of the free energy landscape is not clear.\\
All the explanations based on the free energy landscape
remain often at a qualitative level, because in general
 this landscape cannot be computed and studied exactly.
Only for mean field frustrated systems this ``Landscape Paradigm'' 
\cite{landscape} has
 received a firm theoretical basis. In this case an analytic solution
 of the thermodynamics \cite{beyond} and of the asymptotic out of
equilibrium dynamics \cite{rivista} is available. For these systems
 the free energy landscape can be computed 
\cite{TAP,ptap,crisantitap,tapstatiche,Plefka,Antoine}
 and the partition function, and therefore the equilibrium properties, 
can be recovered as a sum over the free energy minima 
weighted with the Boltzmann factor \cite{young}. 
This approach to the thermodynamics of mean field
 spin glasses is called the TAP approach, because it was introduced
 by Thouless, Anderson and Palmer for the Sherrington-Kirkpatrick model 
\cite{SK}.
In this paper we focus on the class of spin glass models which
reproduce phenomenologically some features of structural glasses
 \cite{vetri1,mcep-spin}.
To understand the relationship between free energy landscape and 
dynamical behaviour we generalise the TAP approach to dynamics. 

\section{The TAP approach}\label{}
\subsection{Static TAP equations}\label{tapstat}
In the following we show 
how the free energy landscape, also called TAP free energy,
can be derived for the p-spin spherical model \cite{ptap,crisantitap}. 
The aim of this section is to present in a simple case the strategy 
which we have followed \cite{tapg} to compute the dynamical TAP
equations.\\
The p-spin Hamiltonian reads:
\begin{equation}\label{hp-spin}
H(\{S_{i}\})=-\sum_{1\leq i_{1}<\cdots <i_{p}\leq N} J_{i_{1},\cdots,i_{p}}
S_{i_{1}}\cdots S_{i_{p}}\quad ,
\end{equation} 
where the couplings are Gaussian variables with zero mean and average
 $\overline{J_{i_{1},\cdots,i_{p}}^{2}}=\frac{p!}
{2N^{p-1}}$.
The TAP free energy 
$\Gamma (\beta , m_{i},l)$, which depends on the 
magnetisation $m_{i}$ at each site $i$ and on the spherical
 parameter $l$, is the Legendre transform of the
``true'' free energy:    
\begin{equation}\label{gdef}
-\beta \Gamma (\beta , m_{i},l)=
\ln  \int_{-\infty}^{+\infty} \prod_{i=1}^{N} dS_{i}  \exp \left( 
-\beta H(\{S_{i}\})-\sum_{i} h_{i} (S_{i}-m_{i})- \frac{\lambda}{2} 
\sum_{i=1}^{N} (S_{i}^2-l)\right)\quad .
\end{equation} 
The Lagrange multipliers $h_{i}(\beta)$ fix the magnetisation at each site $i$:
 $\left<S_{i}\right>=m_{i}$ and $\lambda(\beta)$ enforces the condition
$\sum_{i=1}^{N} \left<S_{i}^2-l\right>=0$. $\left< \cdot \right>$ denotes the thermal average and $N$ is the number of spins. \\
Once $\Gamma $ is known, the equation $-\frac{2}{N} 
\frac{\partial \beta \Gamma}{\partial l} {\Big \vert} _{l=1}=
\lambda$ fixes the spherical constraint ($\sum_{i}S_{i}^{2}=N$) and 
gives the spherical multiplier as a function of $m_{i}$, whereas 
$-\frac{\partial \beta \Gamma}{\partial m_{i}} {\Big \vert} _{l=1}=
h_{i}$ are the TAP equations, which fix the values of local magnetisations.\\
The standard perturbation expansion for the generalised potential $\Gamma $
 is rather involved \cite{doubletransform,crisantitap} and cannot be
 directly applied to the Ising case. Thus, we prefer to follow 
the approach developed for the 
Sherrington-Kirkpatrick model by T. Plefka 
\cite{Plefka} and A. Georges and S. Yedidia \cite{Antoine} 
because it is simple and can be
directly applied to all mean field spin glass models. 
They obtained the TAP free energy for the Sherrington-Kirkpatrick model 
expanding
 $-\beta \Gamma$ in powers of
 $\beta$ around $\beta =0$.
For a general system this corresponds to a 
$\frac{1}{d}$ expansion ($d$ being the spatial dimension) around mean
 field theory \cite{Antoine}; so it is not surprising that for mean 
field spin glass models only a finite number of terms survives. The 
zeroth- and first-order terms give the ``na{\"\i}ve'' TAP free energy, whereas
 the second term is the Onsager reaction term.

From the definition of $-\beta \Gamma$ given in equation (\ref{gdef}), 
we find\footnote{We are neglecting a useless additive constant in
 $\Gamma$. A term in $\Gamma$, that does not depend 
on $l$ and $m_{i}$, has no influence on thermodynamics.} that the
zeroth-order term is the entropy of non interacting 
spherical spins constrained to have magnetisation $m_{i}$:
\begin{equation}\label{g0sferi}
-\beta \Gamma(\beta, m_{i},l) {\Bigg \vert}_{\beta =0}=\frac{N}{2}
\ln \left( l-\frac{1}{N}\sum_{i=1}^{N} m_{i}^2 \right)\quad .
\end{equation}
Using the Lagrange conditions and that the spins are decoupled at $\beta =0$
 we find that the linear term in the power expansion of the
 TAP free energy equals:
\begin{equation}\label{g1a}
-\beta \frac{\partial (\beta\Gamma)}{\partial \beta} {\Bigg \vert}
_{\beta =0}=\beta \sum_{1\leq i_{1}<\cdots <i_{p}\leq N} J_{i_{1}
,\cdots,i_{p}} m_{i_{1}}\cdots m_{i_{p}} \quad.
\end{equation}
This ``mean field'' energy together with the zeroth-order term
 gives the standard mean field theory, which becomes exact for infinite-ranged
ferromagnetic system. The Onsager reaction term comes from the second
 derivative of $\Gamma $:
\begin{equation}\label{g2a}
 -\frac{\beta ^{2}}{2} \frac{\partial ^{2}(\beta\Gamma)}{\partial 
\beta ^{2}} {\Bigg \vert}
_{\beta =0}  =  \frac{\beta ^{2}}{2}\left< \left(\sum_{1\leq i_{1}<\cdots
<i_{p}\leq N} Y_{i_{1},\cdots,i_{p}}\right)^{2}\right>^{c}_{\beta =0}\quad ,\\
\end{equation}
\[
Y_{i_{1},\cdots,i_{p}}=  J_{i_{1},\cdots,i_{p}}S_{i_{1}}\cdots S_{i_{p}}-
(S_{i_{1}}-m_{i_{1}})m_{i_{2}}\cdots m_{i_{p}}-
\]
\[
\cdots -m_{i_{1}}\cdots 
m_{i_{p-1}}(S_{i_{p}}-m_{i_{p}})\quad .
\]
To compute (\ref{g2a}) we have used the following Maxwell relations:
\begin{eqnarray}\label{maxwell}
\frac{\partial h_{i}}{\partial \beta}{\Bigg \vert}_{\beta =0} & = & 
-\frac{\partial}
{\partial m_{i}}\frac{\partial (\beta\Gamma)}{\partial \beta}
{\Bigg \vert}_{\beta =0}\\
\frac{\partial \lambda}{\partial \beta}{\Bigg \vert}_{\beta =0} & = & 
-\frac{2}{N}\frac{\partial}
{\partial l}\frac{\partial (\beta\Gamma)}{\partial \beta}
{\Bigg \vert}_{\beta =0}
\end{eqnarray}
Using the statistical properties of the couplings it is easy to check
 that the only terms giving a contribution of the order of $N$ correspond
 to the squares of $J_{i_{1},\cdots,i_{p}}$:
\begin{equation}\label{g2b}
 -\frac{\beta ^{2}}{2} \frac{\partial ^{2}(\beta\Gamma)}{\partial 
\beta ^{2}} {\Bigg \vert}
_{\beta =0}  =  \frac{\beta ^{2}}{2}\sum_{1\leq i_{1}<\cdots <i_{p}\leq N} 
\left<Y^{2}_{i_{1},\cdots,i_{p}}\right>^{c}_{\beta =0} \quad .
\end{equation}
Using again the statistical properties of the couplings and neglecting
 terms giving a contribution of an order smaller than $N$ we find that the 
reaction term depends on $m_{i}$ through the overlap $q=\frac{1}{N}\sum_{i}m_{i}^{2}$ only:
\begin{equation}\label{g2c1}
 -\frac{\beta ^{2}}{2} \frac{\partial ^{2}(\beta\Gamma)}{\partial 
\beta ^{2}} {\Bigg \vert}
_{\beta =0} =  \frac{\beta ^{2} N}{4}\left(l^p-q^p-p(lq^{p-1}-q^{p})
\right)\quad .
\end{equation}
Higher derivatives lead to terms which can
 be neglected because they are not of the 
order of N \cite{crisantitap,Antoine};
 so collecting (\ref{g0sferi}),(\ref{g1a}) and (\ref{g2c1})
 we find the TAP free energy for spherical p-spin models. 
Differentiating the free energy with respect to magnetisations $m_{i}$ and the
 spherical parameter $l$ one finds the
 TAP equations. 
These equations admit for certain temperatures an infinite
 number of solutions. This is a fundamental characteristic
 and difficulty of mean field spin glasses.\\
It has been shown \cite{young,crisantitap} that 
the weighted sum of the local
 minima of the TAP free energy gives back equilibrium
 results found by the replica or the cavity method \cite{beyond}:
$Z=\sum_{\alpha }e^{-N \beta  f_{\alpha }}$, 
where $f_{\alpha }$ is the TAP free energy of a stable
 solution $\{m_{i}^{\alpha } \}$ of TAP equations. 
Note that states which do not have the minimum free energy can dominate
 the previous sum if their number is very large.

Let us conclude this section with few comments 
on the derivation of the static TAP equations. 
First of all we remark that we have improperly 
called  $\Gamma(\beta ,m_{i},l)$ a Legendre transform. 
Indeed the function $\Gamma(\beta ,m_{i},l)$
is the generating functional of proper vertices \cite{zj}. 
This function may have many minima and is not convex in general.
Finally we want to point out a striking difference, which arises
 in the computation of $\Gamma $, between completely 
connected and finite connectivity mean field models. For the former 
the expansion of $\Gamma $ in powers of $\beta $ stops at the second order
 in $\beta $. Whereas for the latter the expansion contains all 
the powers of $\beta $. 
Roughly speaking for the Sherrington-Kirkpatrick model the
 only non trivial term 
in $\Gamma $ is the reaction term, which represents the 
contribution to the effective field of the {\em i}th spin due to the
influence of the {\em i}th spin on the others. Whereas for its counterpart
on a Bethe lattice \cite{boman,Antoine} the interaction 
between two neighbouring spins has to be taken into 
account exactly, i.e. one has to take into account not only
 the reaction of the neighbours of $S_{i}$ due to the presence of $S_{i}$,  
 but also the reaction of the reaction and so on. 
\label{sec:section2}
\subsection{Dynamical TAP equations}
In the following we focus on a Langevin relaxation dynamics 
for mean field glassy systems. Standard field theoretical 
manipulations \cite{zj} lead to the Martin-Siggia-Rose 
generating functional for the expectation values of $s_{i}(t)$.\\
Within the superspace notation \cite{zj,kurfranz} the 
dynamics and the static theory are formally very similar \cite{kurfranz}. 
As a consequence dynamical TAP equations 
can be derived  straightforwardly generalising the method
described in the previous section. We refer to \cite{tapg} for a detailed
 derivation. Once the dynamical TAP free energy is known, the dynamical TAP 
equation are obtained from the Lagrange relation for the supermagnetisation. 
In the following we simply quote
the result \cite{tapg}:
\begin{eqnarray}\label{tc}
\frac{\partial}{\partial t}{\Big (} C(t,t')-Q(t,t'){\Big )} &=&2R(t',t)-\lambda(t)
{\Big (}C(t,t')-Q(t,t'){\Big )}
+\mu \int_{0}^{t'}dt''{\Big (}C(t,t'')^{p-1}-Q(t,t'')^{p-1}{\Big )}
R(t',t'')\nonumber\\
&+& \mu (p-1)\int_{0}^{t}dt''{\Big (}C(t'',t')-Q(t'',t'){\Big )}R(t,t'')C(t,t'')^{p-2}\quad ,\\
\nonumber\\
\label{tr}
\frac{\partial}{\partial t}R(t,t')&=&-\lambda(t)R(t,t')+\delta(t-t')
+\mu (p-1)\int _{t'}^{t} dt''R(t,t'')R(t'',t')C(t,t'')^{p-2}\quad ,\\
\nonumber\\
\label{tm}
\left(\frac{\partial}{\partial t}+\lambda (t)\right)m_{i}(t)&=&
\beta h_{i}(t)+\beta 
\sum_{1\leq i_{2}< \dots <i_{p}\leq  N }'J_{i,i_{2},\dots ,i_{p}}
m_{i_{2}}(t)\cdots m_{i_{p}}(t)\nonumber \\
&+&\mu (p-1) \int_{0}^{t} dt'' {\Big (}C(t,t'')^{p-2}-Q(t,t'')^{p-2}){\Big )}
R(t,t'')m_{i}(t'')\quad ,
\end{eqnarray}
where $C(t,t')=\frac{1}{N}\sum_{i=1}^{N}\left<s_{i}(t)s_{i}(t')\right>$
 is the correlation function, $R(t,t')=\frac{1}{N}
\sum_{i=1}^{N}\frac{\partial\left< s_{i}(t)\right>}{\partial h_{i}(t')}$ is the
 response function to magnetic fields $h_{i}(t)$ coupled to the
 spins $S_{i}$, $Q(t,t')=\frac{1}{N}\sum_{i=1}^{N}m_{i}(t)m_{i}(t')$ 
is the overlap 
function, $m_{i}(t)$ are the local magnetisation, $\mu =p\beta ^{2}/2$ and 
$\lambda (t)$ is the spherical constraint which fixes $C(t,t)=1$.
 The correlation function satisfies the boundary 
condition $C(t,0)=Q(t,0)$ and 
 magnetisations fulfil the initial
 conditions $m_{i}(0)=s_{i}^{0}$ .
 Note that now $\left< \cdot \right>$ means the average
 over the thermal noise.\\
 Moreover the spherical
 condition $C(t,t)=1$ fixes $\lambda $ as a function of time through the
 equation:
\begin{eqnarray}\label{lambda}
\lambda (t){\Big (}1-q(t){\Big )}&=&1+\frac{1}{2}\frac{dq}{dt}+\mu 
\int_{0}^{t}dt''{\Big (}C(t,t'')^{p-1}-Q(t,t'')^{p-1}{\Big )}R(t,t'')
\nonumber\\
&+&\mu (p-1) \int_{0}^{t}dt''{\Big (}C(t'',t)-Q(t'',t){\Big )}R(t,t'')C(t,t'')^{p-2}\quad ,
\end{eqnarray}
where $q(t)=Q(t,t)$.\\
Three important remarks are in order on these equations.
First of all if one takes for initial condition  
a uniform average  
over all possible configurations as in \cite{leticia},
 then the magnetisations are equal to zero at $t=0$ and there 
is no boundary condition on the correlation function \cite{tapg}. 
In this case we find that the equation (\ref{tm}) is trivially satisfied and
 equations 
(\ref{tc}), (\ref{tr}) and (\ref{lambda}) reduce 
to the ones considered in \cite{leticia}. 
Moreover we notice that in the zero temperature limit the equation (\ref{tm})
 coincides with a simple gradient descent, as should be when the 
 thermal noise is absent. 
Finally, it is interesting to remark that the equations on 
local magnetisations do not
 have at finite times 
the form of a gradient descent in the free energy landscape since
 the Onsager reaction term is non-Markovian.
 This is natural because it represents the contribution 
to the effective field of the $i$th spin 
 due to the influence at previous times 
of the $i$th spin on the others.
\subsection{Asymptotic analysis}\label{an}
In the following we perform an asymptotic analysis of the equations
 (\ref{tc}), (\ref{tr}), (\ref{tm}) and (\ref{lambda}). For the sake of
 simplicity we will take $h_{i}(t)=0$ in (\ref{tm}). \\
Two asymptotic behaviour have been found for the p-spin spherical model
depending on the choice of the initial conditions \cite{leticia,Alain,Franz}:
\begin{itemize}
\item True ergodicity breaking: the system equilibrates 
 in a separate ergodic component. Asymptotically time homogeneity 
 and fluctuation-dissipation theorem (FDT) hold \cite{Alain,Franz}.
In this case, following \cite{Alain,Franz}, we take for 
the asymptotic form of the two time quantities the Ansatz:
\begin{eqnarray}\label{eqansatz}
C(t,t')=C_{FDT}(t-t')&\qquad ,\qquad &R(t,t')=R_{FDT}(t-t')\\
R_{FDT}(\tau )=-\theta(\tau )\frac{dC_{FDT}(\tau )}{d\tau }&\qquad ,\qquad &
Q(t,t')=q\\
\lim_{\tau \rightarrow \infty }C_{FDT}(\tau )=q.&\qquad \qquad  &
\end{eqnarray} 
\item Slow dynamics: the system does not equilibrate.
 Asymptotically two time sectors can be identified. 
 In the first one (FDT regime),
 which corresponds to finite time differences $|t-t'| \sim O(1)$, 
 ($t>>1$, $t'>>1$),
 the system has a pseudo-equilibrium dynamics since
 FDT and time translation invariance hold asymptotically. 
In the second one (ageing regime), which corresponds
 to ``infinite'' time differences $|t-t'|\sim t'$, 
FDT and time translation invariance do not apply and the system
 ages \cite{leticia}. In this case, following \cite{leticia}, 
we take for finite time separations the
 Ansatz corresponding to equilibrium dynamics, but with $Q(t',t)=q'$.
Whereas
 for the ageing sector we take the Ansatz\footnote{The asymptotic
 equations are obtained neglecting the time derivatives. This has as
 a consequence that from an asymptotic solution we obtain infinitely 
 many others
 by re-parameterisation \cite{leticia}. For the sake of clarity 
in the following we focus on the particular
 parameterisation shown in equations (\ref{slansatz1}), (\ref{slansatz2})
 and (\ref{slansatz3}) . } \cite{leticia}:
\begin{eqnarray}\label{slansatz1}
C(t,t')=q C_{ag}(\lambda )&\qquad ,\qquad &t R(t,t')=R_{ag}(\lambda )\\
R_{ag}(\lambda )= x q \frac{d C}{d \lambda }&\qquad ,\qquad &
Q(t,t')=q' Q_{ag}(\lambda )\label{slansatz2}\\
C_{ag}(1)=Q_{ag}(1)=1 &\qquad ,\qquad & \lambda =\frac{t'}{t},\label{slansatz3}
\end{eqnarray}
where $x$ parameterises the violation of FDT \cite{leticia}. 
\end{itemize}
The asymptotic solutions arising from the previous Ans{\"a}tze can be grouped in 
three classes.
\subsubsection{Equilibrium dynamics.}\label{ed}
 We denote respectively by 
$\lambda ^{\infty }$ and $m_{i}^{\infty }$ the
 asymptotic values of the spherical multiplier and 
of the local magnetisations. Plugging the equilibrium dynamics Ansatz
 into the dynamical TAP equations
 we find that the equations on $m_{i}^{\infty }$ and 
$\lambda ^{\infty }$ are 
the corresponding static TAP equations.
 In the asymptotic limit the equations (\ref{tc}) and (\ref{tr}) 
on the correlation
 and the response functions reduce to: 
\begin{equation}\label{eqtapc}
 \left(\frac{d}{d \tau }+\lambda ^{\infty }-\mu \right)C(\tau)+
\mu +1-\lambda ^{\infty }=-\mu \int_{0}^{\tau }d\tau ' C(\tau  -\tau ')^{p-1}
 \frac{dC(\tau ')}{d\tau '}\quad .
\end{equation}
The above equation describes the equilibrium dynamics inside the ergodic 
component associated to a TAP solution $\{m_{i}^{\infty} \}$. Note
 that this asymptotic dynamical 
solution is consistent with the assumption of an 
equilibrium dynamics 
 only if $\{m_{i}^{\infty} \}$ is 
 a local minimum of the free energy. \\
Since this asymptotic solution represents the
equilibration in a stable TAP state $\{m_{i}^{\infty} \}$, it
 is quite natural to associate to this solution
 an initial condition belonging to this state.
This interpretation is suggested by the results of \cite{Alain,Franz}.
 Indeed in \cite{Alain,Franz} 
the low temperature dynamics has been studied 
 starting from an initial condition belonging to the TAP states which are the 
equilibrium states at a temperature $T'$. 
In \cite{Alain,Franz} it has been shown
 that the system relaxes in the 
 TAP states associated to the initial condition. It is easy to show
 that the equation 
 satisfied by $C(\tau )$ in \cite{Alain,Franz} can be written in the form
(\ref{eqtapc}).\\
Moreover it is interesting to note that the equations (\ref{tm}) 
on local magnetisations reduce in the long-time limit
to a gradient descent in the free energy landscape with an extra term
which vanishes at large time.
\subsubsection{Weak ergodicity breaking.}\label{wed}
The asymptotic analysis in the time sector corresponding to finite 
 time differences leads to the same equation (\ref{eqtapc}) 
 for the correlation and the response functions. Whereas for infinite time
 differences we find 
  that the asymptotic equations admit the solution: $q'=0$, $q$ which verifies
 the equation of the overlap of the threshold states 
\cite{leticia,ptap,crisantitap},
 $x=\frac{(p-2)(1-q)}{q}$ and $C_{ag}(\lambda)$ and $R_{ag}(\lambda)$, which
 satisfy the same equations found in \cite{leticia}.
The equation (\ref{lambda}) on the spherical multiplier reduces to:
 $\lambda ^{\infty }=(1-q)^{-1}+\mu (1-q^{p-1})$ 
 and the asymptotic value of the local magnetisations 
 $m_{i}^{\infty }$ is zero. This is exactly the same asymptotic solution
 found in \cite{leticia} for random initial conditions. 
 Therefore it is natural to associate to this solution a random initial
 condition, which is not correlated with any particular stable TAP
 state.\\
 Note that the difference between $q$ and $q'$ 
 clearly marks that the system does not 
 equilibrate in a single ergodic component.
\subsubsection{Between true and weak ergodicity breaking.}\label{twed}
In the following we consider the asymptotic solution which corresponds to
 slow dynamics with $q=q'$. In this case we find the same solution of 
section II.C.2 except that $q'=q$ and $Q_{ag}(\lambda )=C_{ag}(\lambda )$.
As a consequence the local magnetisations do not vanish in the long-time 
limit. These results indicate that at very large times  
the system has almost thermalized within 
 a threshold 
 state. Anyway the slow behaviour of the overlap function $Q(t,t')$
 implies that the local magnetisations evolve forever, even if more
 and more slowly. In other words if one waits a time $t_{w}(>>1)$ the systems
 seems to be equilibrated in a certain threshold states on timescales
 $\Delta t <<t_{w}$; however on timescales of the same order of $t_{w}$ 
 the system continue to evolve.\\
To understand the slow evolution of $m_{i}(t)$ it 
is important to recall that the threshold states are characterised
 by a spectrum of the free energy Hessian which is a semicircle law with minimum eigenvalue
 equal to zero \cite{ptap}. 
As a consequence the free energy landscape around 
threshold states is characterised by
 almost flat directions. 
At large times, the equations satisfied by $m_{i}(t)$ corresponds to 
a gradient descent in the free energy landscape with an extra term
 which vanish in the long-time limit. Because of almost 
 flat directions this vanishing term plays
 a fundamental role and is responsible for ageing. 
 In fact at large times the dynamics takes place only along 
 almost flat directions and this vanishing function of time acts as 
 a vanishing source of drift, so the larger is the time, the
 weaker is the drift and the slower is the evolution: the system
 ages. \\
Finally we remark that it seems natural that the initial conditions 
related to this asymptotic solution 
are the configurations typically reached in the long-time dynamics
 (starting from a random initial condition).
In fact a way to obtain this
 asymptotic solution starting from a random initial condition 
is to introduce fields $h_{i}(t)$ which enforce the
 condition $\lim_{t\rightarrow \infty }1/N\sum_{i=1}^{N}m_{i}(t)^{2}=q'
=q_{th}$ (where $q_{th}$ is the overlap of threshold states
\cite{leticia,ptap,crisantitap}).
There are many different way to fix the fields $h_{i}(t)$ 
to enforce this condition; however for each realization of $h_{i}(t)$ 
it is clear that 
 $lim_{t\rightarrow \infty }h_{i}(t)=0$ because 
 the equality between $q'$ and $q_{th}$ is automatically verified in the
 long-time limit.
 The vanishing of the local magnetisations is due 
\cite{Alain2} to 
 the many possible channels that the system can follow in the energy landscape.
 The role of magnetic fields $h_{i}(t)$ 
is to bring the system along one of the possible channel.
\section{Free energy landscape and long-time dynamics}
\label{sec:section3}
At finite times, the dynamics cannot be represented as 
an evolution in the free energy landscape because the Onsager reaction term
 in (\ref{tm}) is non-Markovian. However in the long time regime 
a connection between the free energy 
landscape and the dynamical evolution can be established.\\
For initial conditions leading to an equilibrium dynamics,
 i.e. the equilibration in a stable TAP state $\{m_{i}^{\infty }\}$,
 the equations on the local magnetisations imply that the relaxation
 of $\{ m_{i}(t) \}$ toward $\{m_{i}^{\infty }\}$ coincides
  with a gradient descent in the free energy
 landscape  with an extra term
 going to zero at large times.\\
Conversely, in the most interesting and the most physical
 case of random initial conditions (corresponding to a quench from
 infinite temperature)
the local magnetisations vanish at large times.
Anyway a description of the asymptotic dynamics as an evolution
 in the free energy landscape makes sense
 also in this case. The local magnetisations vanish asymptotically
 because the dynamical probability measure at large time tends toward
 a static probability measure which is broken in separate ergodic components,
 i.e. the threshold states. One can think at the probability density in
 configuration space as a wave packet which breaks continuously in
 sub-packets. Within this picture, the dynamical evolution is characterised
 by two effects: the cloning \cite{Alain2} of each packet 
in sub-packets and the slow motion of each single packet. 
To avoid the spreading of the dynamical measure and to capture only 
the slow motion, one can take for initial condition a configuration 
typically reached in the long-time dynamics (starting from random initial
 conditions). This procedure leads to the asymptotic solution analysed in
 section II.C.3, in which the correlation and the response functions 
have the same asymptotic behaviour that for a random 
initial condition. Moreover
 $C(t,t')$ and $Q(t,t')$ are equal in the ageing time regime.
 Thus, also the ageing dynamics obtained
 starting from a random initial condition 
can be represented in terms of the equation
 on $m_{i}(t)$, i.e. as a motion in 
the flat directions of the free energy landscape.

In conclusion, through the dynamical TAP approach we have shown that 
the long-time dynamics can be represented as
a gradient descent in the free energy landscape
with an extra term going to zero at large time. This result allow one to make
a straightforward connection between static free energy landscape 
and long-time dynamics and to give to the former a meaningful dynamical
interpretation. 
In fact, consider all the stationary and stable
dynamical probability distributions $P_{\alpha }(\{s_{i}(t) \})$. We have 
shown that the  
local magnetisations $m_{i}^{\alpha }=\left< s_{i} \right> _{\alpha }$ calculated with
 the probability law $P_{\alpha }(\{s_{i}(t) \})$ are the local minima 
of the TAP free energy. This gives to static
 TAP solutions a dynamical interpretation 
 in which the properties of stationarity and stability in
 the free energy landscape are directly related to the 
properties of stationarity and stability of dynamical distributions 
$P_{\alpha }(\{s_{i}(t) \})$. Moreover the relationship, that we have 
 elucidated, between ageing and flat directions in the free energy
 landscape allows one to clarify the important role played
 by the threshold states in the slow dynamics: they are stable states
 having flat directions in the free energy landscape and as a consequence
 they are related to ageing dynamics.
What is missing to a complete dynamical interpretation 
of the free energy landscape is the comprehension of the role played
 by the free energy barriers in the activated dynamics, i.e. to go
 below the threshold energy starting from random initial conditions. 
 Recent progress in this direction has been done in \cite{ioffe}.

\section{Conclusions}
\label{sec:conclusions}
In summary we have found that for the p-spin spherical model 
the representation of the long-time
dynamics as an evolution in the free energy landscape is
correct. This
evolution consists in a gradient descent in the free energy landscape
with an extra term
going to zero at large time. This vanishing source of drift depends
on the history of the system and is crucial for slow dynamics. 
Our results explicitly show that the scenario for slow dynamics found
at zero temperature \cite{laloux} remains valid also at finite temperature:
ageing is due to the motion in the flat directions of the free energy
landscape in presence of a vanishing source of drift.\\
Finally, the relationship between long-time dynamical behaviour and
 local properties of the free energy landscape, which was already found
 in \cite{leticia,Alain,Franz,glauber}, shows up explicitly 
by the study of the dynamical TAP equations. This relationship is very 
important not only from a theoretical point of view, but also
from a technical one. Indeed it allows to obtain information about
the long-time dynamics by a pure static computation \cite{Franz,Remi}.
For these reasons it would be very interesting to generalise the
 study performed in this article to finite dimensional systems. 
In this case the free energy
 landscape cannot be computed exactly and the long-time dynamics cannot
 be solved; however, the formal analogies (due to superspace notation) between
 static and dynamic theory let us hope that one can obtain
 results on the relationship between long-time
 dynamics and free energy landscape only using the symmetry properties
 of the asymptotic solution \cite{leticia2}.
\section{Acknowledgements}
\label{acknowledgements}
I am deeply indebted to L. F. Cugliandolo,
 J. Kurchan and R. Monasson for
numerous, helpful and thorough discussions on this work. I wish also to thank 
 S. Franz and M. A. Virasoro 
for many interesting discussions on the asymptotic
 solution analysed in section 2.3.3. I am particularly 
 grateful to R. Monasson for his constant support and for 
 a critical reading of the manuscript.

\end{document}